\documentclass[preprint,amsmath,amssymb,aps,prd,floatfix,superscriptaddress,nofootinbib]{revtex4-1}
\usepackage{physics}

\usepackage{amsmath,amssymb}
\usepackage{graphicx,subfigure}
\usepackage{color,multirow}
\usepackage[colorlinks,linkcolor=magenta,anchorcolor=cyan,citecolor=blue,plainpages=false]{hyperref}

\hypersetup{colorlinks=true,
    breaklinks=true,
    pdfstartview=Fit,
    linkcolor=blue,
    citecolor=blue,
    urlcolor=blue}

\def\be{\begin{equation}}
    \def\ee{\end{equation}}
\def\ba{\begin{eqnarray}}
    \def\ea{\end{eqnarray}}

\begin{document}
\title{Dark energy and lensing anomaly in Planck CMB data}

\author{Ze-Yu Peng}
\email{pengzeyu23@mails.ucas.ac.cn} 
\affiliation{School of Physical Sciences, University of Chinese
Academy of Sciences, Beijing 100049, China}
\affiliation{International Centre for Theoretical Physics
Asia-Pacific, University of Chinese Academy of Sciences, 100190
Beijing, China}
\affiliation{School of
Fundamental Physics and Mathematical Sciences, Hangzhou Institute
for Advanced Study, UCAS, Hangzhou 310024, China}

\author{Yun-Song Piao}
\email{yspiao@ucas.ac.cn}
\affiliation{School of Physical Sciences, University of Chinese Academy of Sciences, Beijing 100049, China}
\affiliation{International Centre for Theoretical Physics Asia-Pacific, University of Chinese Academy of Sciences, 100190 Beijing, China}
\affiliation{School of Fundamental Physics and Mathematical Sciences, Hangzhou Institute for Advanced Study, UCAS, Hangzhou 310024, China}
\affiliation{Institute of Theoretical Physics, Chinese Academy of Sciences, P.O. Box 2735, Beijing 100190, China}

\begin{abstract}

In this paper, we investigate the impact of the lensing anomaly in Planck cosmic microwave background (CMB) data on the nature of dark energy (DE). We constrain the state equation ($w_0,w_a$) of DE with the lensing scaling parameter $A_L=1$ and varying $A_L$, using the Planck PR3 and two updated Planck PR4 likelihoods, CamSpec and HiLLiPoP respectively, combined with DESI baryon acoustic oscillation (BAO) and Pantheon+ supernova data. As expected, when $A_L$ is allowed to vary, the evolving DE is not preferred due to the degeneracy between $w_0,w_a$ and $A_L$. In particular, we also consider replacing DESI BAO data with pre-DESI BAO in our analysis, and observe that DESI BAO appears to exacerbate the lensing anomaly, which is caused by the smaller matter density $\Omega_m$ it prefers, however, this effect can be offset by the shifts in $w_0$ and $w_a$ preferring the evolving DE. Our work indicates that the lensing anomaly in Planck data is worth carefully reconsidering when new cosmological survey data is combined with CMB.

\end{abstract}

\maketitle

\newpage

\section{Introduction}\label{sec:intro}
The Baryon Acoustic Oscillation (BAO) measurements from the
first-year observation of Dark Energy Spectroscopic Instrument
(DESI), when combined with Cosmic Microwave Background (CMB) and
Supernova (SN) data, indicate at $>2\sigma$ significance level
that dark energy (DE) is evolving \cite{DESI:2024mwx}. This
finding has significant theoretical implications and may
profoundly impact our understanding of the nature of DE. There has
been extensive discussions and reanalyses of this result
\cite{Tada:2024znt,Gu:2024jhl,Luongo:2024fww,Cortes:2024lgw,Colgain:2024xqj,Carloni:2024zpl,Wang:2024dka,Chan-GyungPark:2024mlx,Park:2024jns,Wang:2024pui,Shlivko:2024llw,Dinda:2024kjf,Seto:2024cgo,Roy:2024kni,Gialamas:2024lyw,Toda:2024ncp,Wang:2024hwd,Orchard:2024bve,Wang:2024sgo,Giare:2024gpk,Giare:2024gpk,Jiang:2024viw,Jiang:2024xnu,Dinda:2024ktd,Alfano:2024jqn,Ghosh:2024kyd,Pedrotti:2024kpn,Pang:2024qyh,RoyChoudhury:2024wri,Giare:2024ocw,Wang:2024tjd,Chan-GyungPark:2024brx,Specogna:2024euz,Pang:2024wul,Colgain:2024mtg},
as well as numerous studies on related theoretical models
\cite{Berghaus:2024kra,Giare:2024smz,Escamilla-Rivera:2024sae,Bhattacharya:2024hep,Heckman:2024apk,Chudaykin:2024gol,Du:2024pai,Li:2024qso,Ye:2024ywg,Sohail:2024oki,Wolf:2024eph,Wolf:2024stt,Alestas:2024eic,Bhattacharya:2024kxp,Ye:2024zpk,Li:2024qus,Akthar:2024tua,daCosta:2024grm,Du:2025iow,Wolf:2023uno}.
Furthermore, this $>2\sigma$ significance level for evolving DE
remains consistent when including the recent DESI full-shape
clustering data \cite{DESI:2024hhd}.

It is important to note that the DESI results rely on the Planck
2018 (PR3) CMB data, which dominates the constraints on most
cosmological parameter in the analysis. However, the Planck 2018
data are known to exhibit some mild issues, such as the lensing
anomaly \cite{Calabrese:2008rt,Planck:2018vyg}, i.e. the excess
lensing effect observed in the CMB power spectra. This anomaly may
have non-trivial impacts on the nature for DE due to their
degenerate effect. Ref.~\cite{Chan-GyungPark:2024brx} investigated
the evolution of DE in light of lensing anomaly using non-DESI
data, see also recent \cite{Chan-GyungPark:2025cri}.


The last Planck data release (PR4), which employ the NPIPE
processing pipeline, is reported with slightly more data, lower
noise, and better consistency between frequency channels
\cite{Planck:2020olo}. Recently updated Planck PR4 likelihoods,
including CamSpec \cite{Rosenberg:2022sdy} and HiLLiPoP
\cite{Tristram:2023haj}, both report a weaker lensing anomaly,
especially the latter which is consistent with $A_L=1$ within
$1\sigma$. It is also found that these PR4-based likelihoods show
better consistency with General Relativity (GR) compared to Planck
2018 \cite{Specogna:2024euz,DESI:2024yrg}, and can relax the upper
limit on the sum of neutrino masses \cite{Allali:2024aiv}.
Therefore, it is also necessary to revisit the DESI results for
evolving DE using these updated Planck likelihoods.

In addition, it is also noteworthy that CMB experiments other than
Planck, such as the Atacama Cosmology Telescope (ACT)
\cite{ACT:2020gnv} and the South Pole Telescope (SPT)
\cite{SPT-3G:2021eoc,SPT-3G:2022hvq}, show no signs of excess
lensing effect. A previous study found that that the non-Planck
CMB data seem not to prefer the evolving DE \cite{Giare:2024ocw}.

In this paper, we aim to investigate the impact of the lensing
anomaly in Planck data on the nature of DE. We constrain the state
equation of DE with the lensing scaling parameter $A_L=1$ and
varying $A_L$, using the Planck PR3 and PR4 likelihoods
respectively.
The paper is outlined as follows. We describe the methodology and
data used in this work in Section~\ref{sec:methods}. We present
our results on DE with $A_L=1$ and varying $A_L$ for different
Planck likelihoods in Section~\ref{sec:results}, and the results
with pre-DESI BAO in Section~\ref{sec:pre-desi}. Based on our
results, we discuss the underlying physics of the correlation
between the evolving DE and the lensing anomaly in
Section~\ref{sec:lensing}. We conclude in
Section~\ref{sec:conclusion}. We also present the results of all relevant parameters in Appendix~\ref{sec:appendix}.

\section{Methodology and Data}\label{sec:methods}
We consider the widely adopted Chevallier-Polarski-Linder (CPL)
parameterisation to account for the evolution of DE, in which the
DE equation of state is:
\begin{equation}
    w(a) = w_0 + w_a(1-a).
\end{equation}
We will refer to the corresponding model as $w_0w_a$CDM (while
$w_0=-1$ and $w_a=0$ corresponds to $\Lambda$CDM), and refer to
the models with $A_L$ varying (so that the lensing spectrum is
rescaled as $C_{\ell}^{\phi \phi} \longrightarrow A_L
C_{\ell}^{\phi \phi}$) as $\Lambda$CDM+$A_L$ and $w_0w_a$CDM+$A_L$
respectively.

\begin{table}[htbp]
    \centering
    \begin{tabular}{|l|c|c|c|c|}
        \hline
        Tracer & $z_{\mathrm{eff}}$ & $D_M/r_d$ & $D_H/r_d$ & $D_V/r_d$ \\
        \hline
        BGS & 0.295 & --- & --- & $7.93\pm0.15$ \\
        LRG1 & 0.510 & $13.62\pm0.25$ & $20.98\pm0.61$ & --- \\
        LRG2 & 0.706 & $16.85\pm0.32$ & $20.08\pm0.60$ & --- \\
        LRG3+ELG1 & 0.930 & $21.71\pm0.28$ & $17.88\pm0.35$ & --- \\
        ELG2 & 1.317 & $27.79\pm0.69$ & $13.82\pm0.42$ & --- \\
        QSO & 1.491 & --- & --- & $26.07\pm0.67$ \\
        Lya-QSO & 2.330 & $39.71\pm0.94$ & $8.52\pm0.17$ & --- \\
        \hline
    \end{tabular}
    \caption{BAO likelihoods from DESI Year 1 measurements \cite{DESI:2024mwx}. }
    \label{tab:bao1}
\end{table}

\begin{table}[htbp]
    \centering
    \begin{tabular}{|l|c|c|c|c|}
        \hline
        Tracer & $z_{\mathrm{eff}}$ & $D_M/r_d$ & $D_H/r_d$ & $D_V/r_d$ \\
        \hline
        6dF & 0.106 & --- & --- & $2.98\pm0.13$ \\
        MGS & 0.15 & --- & --- & $4.47\pm0.17$ \\
        BOSS Galaxy & 0.38 & $10.23\pm0.17$ & $25.00\pm0.76$ & --- \\
        BOSS Galaxy & 0.51 & $13.36\pm0.21$ & $22.33\pm0.58$ & --- \\
        eBOSS LRG & 0.70 & $17.86\pm0.33$ & $19.33\pm0.53$ & --- \\
        eBOSS ELG & 0.85 & --- & --- & $18.33^{+0.57}_{-0.62}$ \\
        eBOSS QSO & 1.48 & $30.69\pm0.80$ & $13.26\pm0.55$ & --- \\
        eBOSS Lya& 2.33 & $37.6\pm1.9$ & $8.93\pm0.28$ & --- \\
        eBOSS Lya-QSO & 2.33 & $37.3\pm1.7$ & $9.08\pm0.34$ & --- \\
        \hline
    \end{tabular}
\caption{BAO likelihoods from pre-DESI measurements, including 6dF
Galaxy Survey \cite{Beutler:2012px}, SDSS DR7 Main Galaxy Sample
(MGS) \cite{Ross:2014qpa}, BOSS DR12 Galaxies \cite{BOSS:2016wmc},
and eBOSS DR16 \cite{eBOSS:2020yzd}. }
    \label{tab:bao2}
\end{table}

The Planck CMB PR3 and PR4 likelihoods are described as follows:
\begin{itemize}
    \item \textbf{Plik}: The Planck 2018 (PR3) \texttt{Plik} likelihood for high-$\ell$ TT, TE, and EE power spectra, the \texttt{Commander} likelihood for low-$\ell$ TT spectrum, and the \texttt{SimALL} likelihood for low-$\ell$ EE spectrum \cite{Planck:2019nip}.
    \item \textbf{CamSpec}: The Planck NPIPE (PR4) \texttt{CamSpec} likelihood for high-$\ell$ TT, TE, and EE power spectra \cite{Rosenberg:2022sdy}, the \texttt{Commander} likelihood for low-$\ell$ TT spectrum, and the \texttt{SimALL} likelihood for low-$\ell$ EE spectrum.
    \item \textbf{HiLLiPoP}: The Planck NPIPE (PR4) \texttt{HiLLiPoP} likelihood for high-$\ell$ TT, TE, and EE power spectra \cite{Tristram:2023haj}, the \texttt{Commander} likelihood for low-$\ell$ TT spectrum, and the NPIPE \texttt{LoLLiPoP} likelihood for low-$\ell$ EE spectrum \cite{Tristram:2020wbi}.
\end{itemize}
These Planck likelihoods are combined with the full DESI BAO
likelihoods \cite{DESI:2024mwx} presented in Table~\ref{tab:bao1},
where $z_{\mathrm{eff}}$ is the effective redshift, $r_d$ is the
sound horizon at the drag epoch, $D_M$ is the comoving distance,
$D_H$ is the Hubble distance, and $D_V$ is the angle-averaged
distance. In Section~\ref{sec:pre-desi}, we also consider
replacing DESI BAO with a pre-DESI BAO likelihood, as presented in
Table~\ref{tab:bao2}. For all our analyses, we always include the
CMB lensing likelihoods from Planck PR4 \cite{Carron:2022eyg} and
ACT DR6 \cite{ACT:2023kun,ACT:2023dou}, as well as the Pantheon+
SN Ia data \cite{Scolnic:2021amr}.

We perform the Markov chain Monte Carlo (MCMC) analysis using
\texttt{Cobaya} \cite{Torrado:2020dgo}. The observables are
computed using the cosmological Boltzmann code \texttt{CLASS}
\cite{Blas:2011rf}. We take our MCMC chains to be converged using
the Gelman-Rubin criterion \cite{Gelman:1992zz} with $R-1<0.02$.
The best-fit parameters and corresponding $\chi^2$ values are
obtained using \texttt{PROSPECT}, which employs a simulated annealing algorithm\footnote{We use the global optimization method in \texttt{PROSPECT}, which we find is much more effective in finding global minima than the minimizers integrated within \texttt{Cobaya}.}.
\cite{Holm:2023uwa}.
To quantify the data for the preference of models , we calculate
the Akaike Information Criterion (AIC) \cite {Akaike:1974vps}
defined as $\mathrm{AIC}=2k+\chi^2$, where $k$ is the number of
free parameters in the model.

We adopt wide, uninformative priors for all relevant parameters,
including the six standard $\Lambda$CDM parameters:
$\ln(10^{10}A_s)$, $n_s$, $H_0$, $\omega_b$,
$\omega_{\mathrm{cdm}}$, and $\tau_{\mathrm{reio}}$, the CPL
parameters: $w_0$ and $w_a$, and the lensing scaling parameter:
$A_L$.


\section{Results }\label{sec:results}


Table~\ref{tab:result} presents the marginalized mean and $68\%$
confidence intervals of relevant parameters for $w_0w_a$CDM,
$\Lambda$CDM+$A_L$, and $w_0w_a$CDM+$A_L$ with different Planck
likelihoods. Table~\ref{tab:result} also presents the
$\Delta\chi^2$ and $\Delta$AIC values for each best-fit model
relative to the best-fit $\Lambda$CDM model using the same
dataset. The results of all relevant parameters including the
best-fit values are shown in Appendix~\ref{sec:appendix}.

Fig.~\ref{fig:w0wa} displays the posterior distributions of
$w_0-w_a$ for $w_0w_a$CDM (left) and $w_0w_a$CDM+$A_L$ (right).
The left panel of Fig.~\ref{fig:Al} shows the 1D posterior
distributions of $A_L$ for $\Lambda$CDM+$A_L$ (solid) and
$w_0w_a$CDM+$A_L$ (dashed), while the right panel shows the
scatter plot of $w_0-w_a$ for $w_0w_a$CDM+$A_L$ obtained with
Plik, with color coding for $A_L$.

\begin{table}[htbp]
    \centering
    \setlength{\tabcolsep}{6pt}
    \begin{tabular}{lccccc}
    \hline

    \hline\hline
    model / dataset                                         & $w_0$              & $w_a$                   & $A_L$                     & $\Delta\chi^2$ & $\Delta$AIC \\ \hline
    $\boldsymbol{w_0w_a}$\textbf{CDM}                     &                    &                         &                           &                &             \\
    Plik                                                  & $-0.831\pm 0.063 $ & $-0.73^{+0.29}_{-0.25}$ & ---                       & $-7.56$        & $-3.56$     \\
    CamSpec                                               & $-0.832\pm 0.063$  & $-0.73^{+0.29}_{-0.25}$ & ---                       & $-7.82$        & $-3.82$     \\
    HiLLiPoP                                              & $-0.842\pm 0.063$  & $-0.64^{+0.30}_{-0.24}$ & ---                       & $-5.91$        & $-1.91$     \\ \hline
    $\boldsymbol{\Lambda}$\textbf{CDM+}$\boldsymbol{A_L}$ &                    &                         &                           &                &             \\
    Plik                                                  & ---                & ---                     & $1.083^{+0.033}_{-0.037}$ & $-6.21$        & $-4.21$     \\
    CamSpec                                               & ---                & ---                     & $1.075\pm 0.034 $         & $-4.24$        & $-2.24$     \\
    HiLLiPoP                                              & ---                & ---                     & $1.062\pm 0.035$          & $-2.63$        & $-0.63$     \\ \hline
    $\boldsymbol{w_0w_a}$\textbf{CDM+}$\boldsymbol{A_L}$  &                    &                         &                           &                &             \\
    Plik                                                  & $-0.859\pm 0.063$  & $-0.51^{+0.32}_{-0.25}$ & $1.066\pm 0.041$          & $-11.26$       & $-5.26$     \\
    CamSpec                                               & $-0.851\pm 0.064$  & $-0.58^{+0.32}_{-0.26}$ & $1.051\pm 0.040$          & $-9.41$        & $-3.41$     \\
    HiLLiPoP                                              & $-0.859\pm 0.064$  & $-0.52^{+0.30}_{-0.26}$ & $1.043\pm 0.040$          & $-7.73$        & $-1.73$     \\
    \hline\hline

    \hline
    \end{tabular}%
\caption{The mean and $68\%$ confidence intervals of relevant
parameters for $w_0w_a$CDM, $\Lambda$CDM+$A_L$, and
$w_0w_a$CDM+$A_L$, as well as the best-fit $\Delta\chi^2$ and
$\Delta$AIC values relative to $\Lambda$CDM. The datasets used are
Planck PR3 and PR4 likelihoods, respectively combined with DESI
BAO, CMB lensing, and Pantheon+.}
    \label{tab:result}
\end{table}

\subsection{Constraints with $A_L=1$}

\begin{figure}[htbp]
    \centering
   \includegraphics[width=0.48\linewidth]{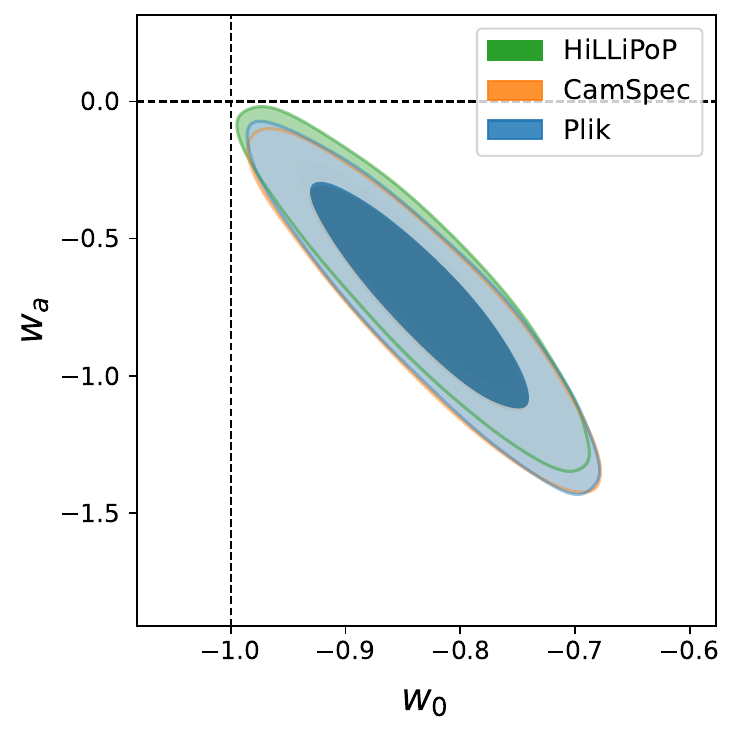}
    \hfill
   \includegraphics[width=0.48\linewidth]{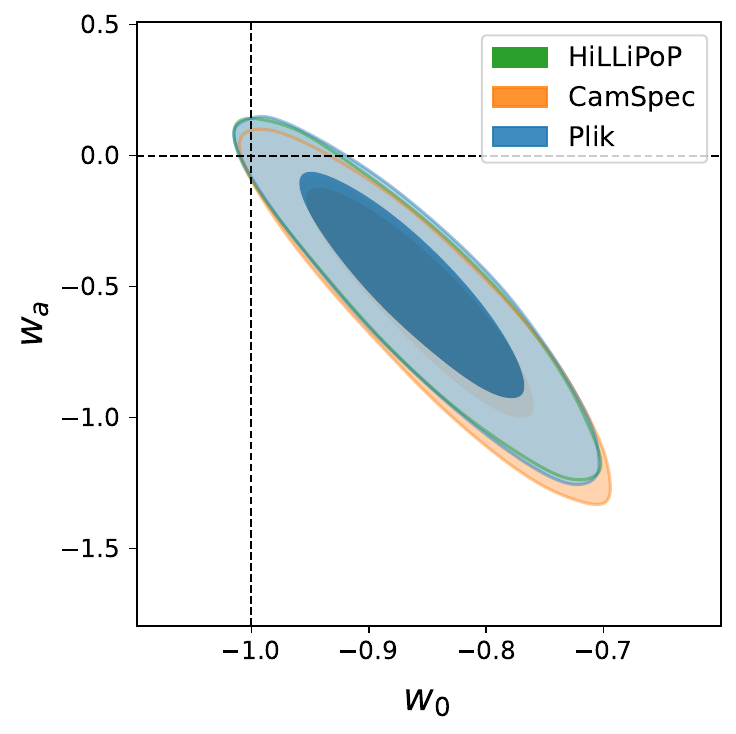}
\caption{Posterior distributions ($68\%$ and $95\%$ confidence
range) of $w_0-w_a$ for $w_0w_a$CDM (left) and $w_0w_a$CDM+$A_L$
(right). The datasets used are Planck PR3 and PR4 likelihoods,
respectively combined with DESI BAO, CMB lensing, and Pantheon+.}
    \label{fig:w0wa}
\end{figure}

In this case, we obtain $w_0=-0.831\pm 0.063$ and
$w_a=-0.73^{+0.29}_{-0.25}$ for Plik, $w_0=-0.832\pm 0.063$ and
$w_a=-0.73^{+0.29}_{-0.25}$ for CamSpec, and $w_0=-0.842\pm 0.063$
and $w_a=-0.64^{+0.30}_{-0.24}$ for HiLLiPoP. As shown in the left
panel of Fig.~\ref{fig:w0wa}, the $w_0-w_a$ posterior obtained
with CamSpec closely matches that obtained with Plik. The
posterior obtained with HiLLiPoP exhibits a slight shift toward
the cosmological constant ($w_0=-1$, $w_a=0$), but is still
consistent with the result with Plik.

The $\Delta\mathrm{AIC}$ values of the best-fit $w_0w_a$CDM model
relative to $\Lambda$CDM are $-3.56$, $-3.82$, and $-1.91$ for
Plik, CamSpec, and HiLLiPoP likelihoods, respectively. In view of
a goodness-of-fit perspective, we once again find that the CamSpec
likelihood maintains the $>2\sigma$ significant level for
$w_0w_a$CDM, while the HiLLiPoP likelihood slightly weakens it,
with $\abs{\Delta\mathrm{AIC}}\lesssim 2$.

\subsection{Constraints with varying $A_L$}

In this case, we obtain $w_0=-0.859\pm 0.063$ and
$w_a=-0.51^{+0.32}_{-0.25}$ for Plik, $w_0=-0.851\pm 0.064$ and
$w_a=-0.58^{+0.32}_{-0.26}$ for CamSpec, and $w_0=-0.859\pm 0.064$
and $w_a=-0.52^{+0.30}_{-0.26}$ for HiLLiPoP. As shown in the
right panel of Fig.~\ref{fig:w0wa}, the $w_0-w_a$ posteriors of
$w_0w_a$CDM+$A_L$ are all consistent with the cosmological
constant within $2\sigma$. The $\Delta\mathrm{AIC}$ values of the
best-fit $w_0w_a$CDM+$A_L$ model relative to $\Lambda$CDM+$A_L$
are $-1.05$, $-1.17$, and $-1.09$ for Plik, CamSpec, and HiLLiPoP
likelihoods, respectively. Therefore, when $A_L$ is allowed to
vary, $w_0w_a$CDM is not preferred.

\begin{figure}[htbp]
    \centering
   \includegraphics[width=0.48\linewidth]{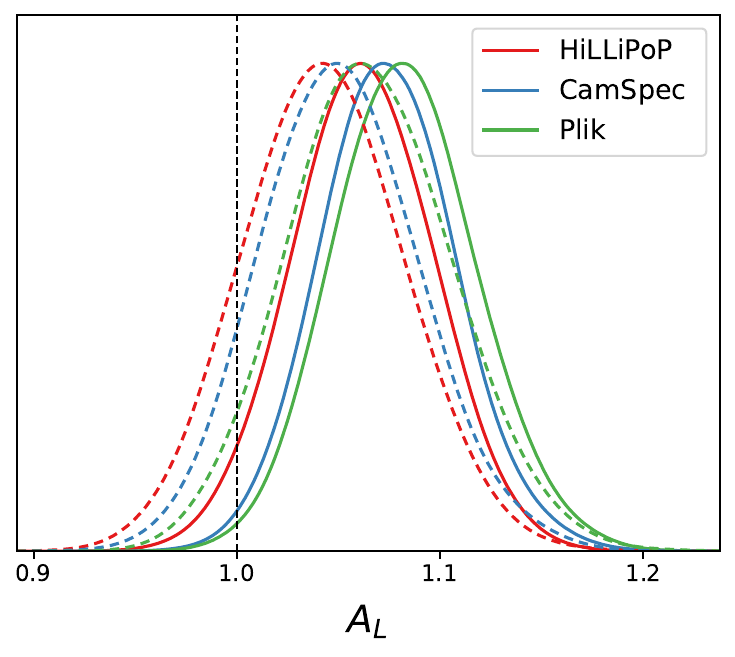}
    \hfill
   \includegraphics[width=0.5\linewidth]{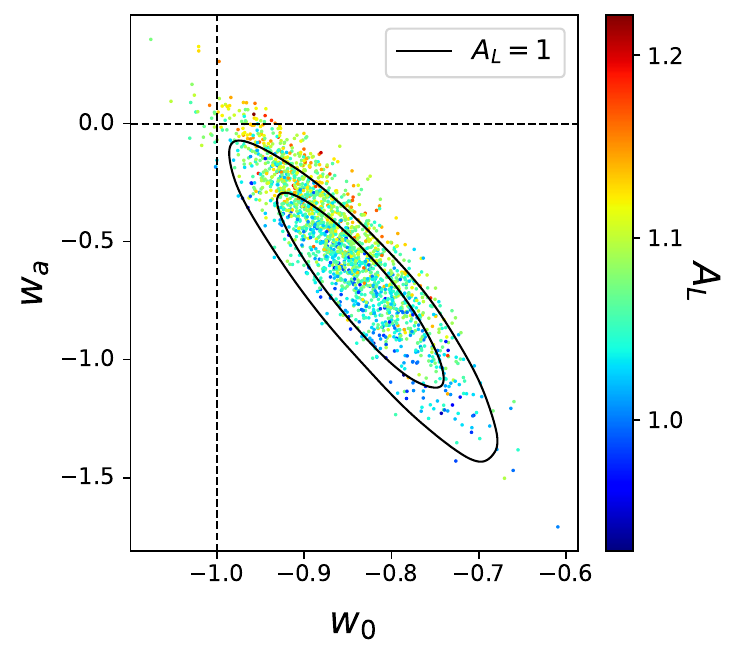}
\caption{\emph{Left panel}: 1D posterior distributions of $A_L$
for $\Lambda$CDM+$A_L$ (solid) and $w_0w_a$CDM+$A_L$ (dashed).
\emph{Right panel}: The scatter plot of $w_0-w_a$ for
$w_0w_a$CDM+$A_L$ (obtained with Plik), with color coding for
$A_L$. The black contour is $w_0w_a$CDM with $A_L=1$ for
comparison.}
    \label{fig:Al}
\end{figure}

It is well known that the Planck data has the lensing anomaly in
$\Lambda$CDM+$A_L$ model. Here, we have
$A_L=1.083^{+0.033}_{-0.037}$ for $\Lambda$CDM+$A_L$ with Plik,
$A_L=1.075\pm 0.034$ for CamSpec, and $A_L=1.062\pm 0.035$ for
HiLLiPoP; while for $w_0w_a$CDM+$A_L$, we have $A_L=1.066\pm
0.041$ for Plik, $A_L=1.051\pm 0.040$ for CamSpec, and
$A_L=1.043\pm 0.040$ for HiLLiPoP. Therefore, the lensing anomaly
in Planck data is weaken for $w_0w_a$CDM (see also the left panel
of Fig.~\ref{fig:Al}).
In addition, it is clearly observed from the right panel of
Fig.~\ref{fig:Al} that the $>2\sigma$ preference for $w_0w_a$CDM
is correlated with the abatement of lensing anomaly.



\section{Comparison with pre-DESI BAO}\label{sec:pre-desi}

\begin{figure}[htbp]
    \centering
   \includegraphics[width=0.7\linewidth]{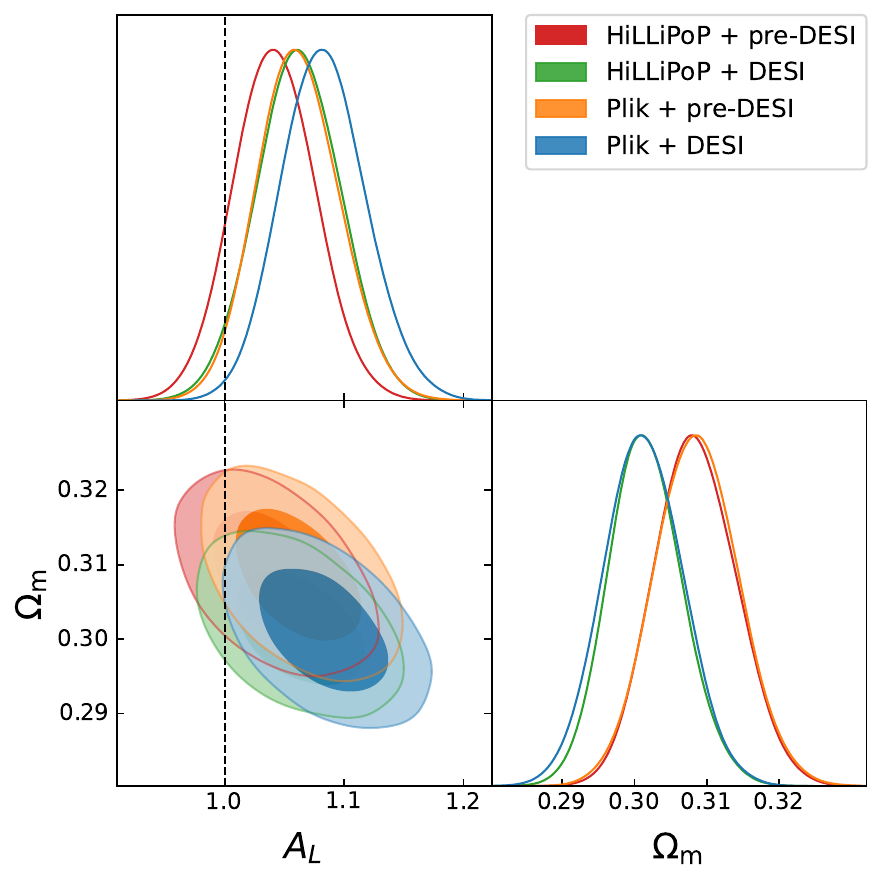}
\caption{1D and 2D marginalized posterior distributions ($68\%$
and $95\%$ confidence range) of $A_L$ and $\Omega_m$ for
$\Lambda$CDM+$A_L$ obtained using Plik and HiLLiPoP, combined with
DESI BAO and pre-DESI BAO respectively.}
    \label{fig:lcdm_pre}
\end{figure}

\begin{figure}[htbp]
    \centering
   \includegraphics[width=0.9\linewidth]{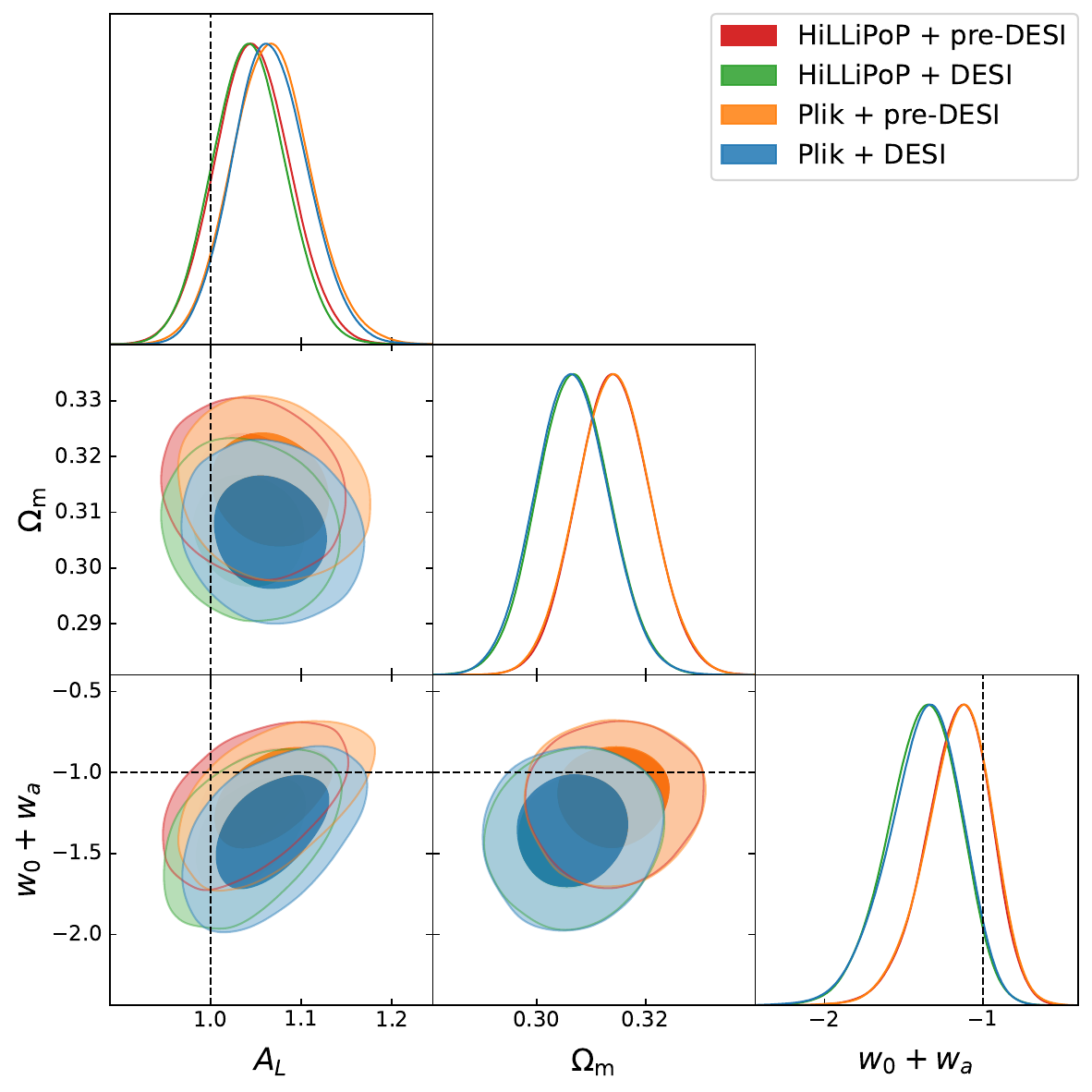}
\caption{1D and 2D marginalized posterior distributions ($68\%$
and $95\%$ confidence range) of $A_L$, $\Omega_m$, and $w_0+w_a$
for $w_0w_a$CDM+$A_L$. The datasets are Planck PR4 likelihoods,
Plik and HiLLiPoP, combined with DESI BAO and pre-DESI BAO
respectively, as well as CMB lensing, and Pantheon+.}
    \label{fig:w0wa_pre}
\end{figure}

In this section, to further clarify the effect of DESI BAO on
lensing anomaly, we consider replacing DESI BAO with pre-DESI BAO
for comparison. Table~\ref{tab:result} presents the marginalized
mean and $68\%$ confidence intervals of relevant parameters for
$\Lambda$CDM+$A_L$ and $w_0w_a$CDM+$A_L$.
Fig.~\ref{fig:lcdm_pre} and Fig.~\ref{fig:w0wa_pre} display the 1D
and 2D posterior distributions of relevant parameters for
$\Lambda$CDM+$A_L$ and $w_0w_a$CDM+$A_L$, respectively.

\begin{table}[htbp]
    \centering
    \setlength{\tabcolsep}{6pt}
    \begin{tabular}{lccccc}
    \hline

    \hline\hline
    model / dataset                                         & $w_0$              & $w_a$                   & $A_L$                   \\ \hline
    $\boldsymbol{\Lambda}$\textbf{CDM+}$\boldsymbol{A_L}$ &                    &                         &                               \\
    Plik                                                  & ---                & ---                     & $1.062\pm 0.035$    \\
    CamSpec                                               & ---                & ---                     & $1.054\pm 0.034 $            \\
    HiLLiPoP                                              & ---                & ---                     & $1.042\pm 0.035$               \\ \hline
    $\boldsymbol{w_0w_a}$\textbf{CDM+}$\boldsymbol{A_L}$  &                    &                         &                                \\
    Plik                                                  & $-0.906\pm 0.062$  & $-0.25^{+0.29}_{-0.23}$ & $1.067\pm 0.044$            \\
    CamSpec                                               & $-0.902\pm 0.063$  & $-0.30^{+0.29}_{-0.24}$ & $1.052\pm 0.041$             \\
    HiLLiPoP                                              & $-0.905\pm 0.062$  & $-0.26^{+0.28}_{-0.23}$ & $1.046\pm 0.042$              \\
    \hline\hline

    \hline
    \end{tabular}%
\caption{The mean and $68\%$ confidence intervals of relevant
parameters for $\Lambda$CDM+$A_L$ and $w_0w_a$CDM+$A_L$. The
datasets are Planck PR3 and PR4 likelihoods, respectively combined
with DESI BAO, CMB lensing, and Pantheon+.}
    \label{tab:pre-result}
\end{table}

The results of $A_L$ in $\Lambda$CDM+$A_L$ with pre-DESI BAO
($A_L=1.062\pm 0.035$ for Plik, $A_L=1.054\pm 0.034$ for CamSpec,
and $A_L=1.042\pm 0.035$ for HiLLiPoP) are smaller than those with
DESI BAO, with similar uncertainties. This indicate that DESI BAO
slightly exacerbates the lensing anomaly in Planck data.

The results of $A_L$ with pre-DESI BAO are similar between
$\Lambda$CDM+$A_L$ and $w_0w_a$CDM+$A_L$, which is different from
those with DESI BAO. This is related to the fact that pre-DESI BAO
itself does not show a preference for evolving DE.

Notebly, as shown in Fig.~\ref{fig:lcdm_pre}, the higher $A_L$
obtained in $\Lambda$CDM+$A_L$ with DESI BAO is correlated with
the smaller $\Omega_m$ preferred by DESI BAO compared to pre-DESI
BAO. This degeneracy exists because a smaller $\Omega_m$ would
suppress the original lensing potential, which should be
compensated by a higher $A_L$. In the $w_0w_a$CDM+$A_L$ model, we
find this degeneracy disappears and there is almost no difference
in the $A_L$ posteriors between DESI BAO and pre-DESI BAO. This is
because the effect of smaller $\Omega_m$ here can be compensated
by the shifts in $w_0$ and $w_a$, as seen in
Fig.~\ref{fig:w0wa_pre}. See also Refs.~\cite{Colgain:2024mtg,Tang:2024lmo} for more discussions on the discrepancies in $\Omega_m$ across different datasets and their effects on evolvong DE.


\section{Discussion on DE and lensing anomaly} \label{sec:lensing}

It is necessary to discuss why the exacerbation of the lensing
anomaly in Planck data for $\Lambda$CDM, caused by DESI BAO, can
be offset by the shifts in $w_0$ and $w_a$. The lensing potential
on the sky (with the comoving distance $\chi$ to the source),
which sets the deflection angle of CMB photons, is given by (in
the Newtonian gauge) \cite{2020moco.book.....D}:
\begin{equation}
    \phi_L(\boldsymbol\theta) = \int_0^\chi \frac{d\chi'}{\chi'}\Phi_W(\boldsymbol{x}(\boldsymbol\theta,\chi'))\qty(1-\frac{\chi'}{\chi}),
\end{equation}
where $\Phi_W\equiv \Psi+\Phi$ is the Weyl potential. Its
evolution in the late universe is determined by the growth function \cite{2020moco.book.....D}
\begin{equation}
    D(a) = \frac{5\Omega_m}{2}\frac{H(a)}{H_0}\int_0^a
    \frac{da'}{\left[a'H(a')/ H_0\right]^3},
\end{equation}
where $H(a)/H_0 =\sqrt{\Omega_m a^{-3} + (1-\Omega_m)
a^{-3(1+w(a))}}$ (assuming the universe is flat). Therefore, a
lower Hubble expanding rate $H(a)$ in the past (corresponding to $w(a)<-1$)
seems to inevitably magnify the Weyl potential, so the lensing
potential,
see also e.g.Refs.~\cite{Green:2024xbb, Loverde:2024nfi}.

\begin{figure}[htbp]
    \centering
   \includegraphics[width=0.49\linewidth]{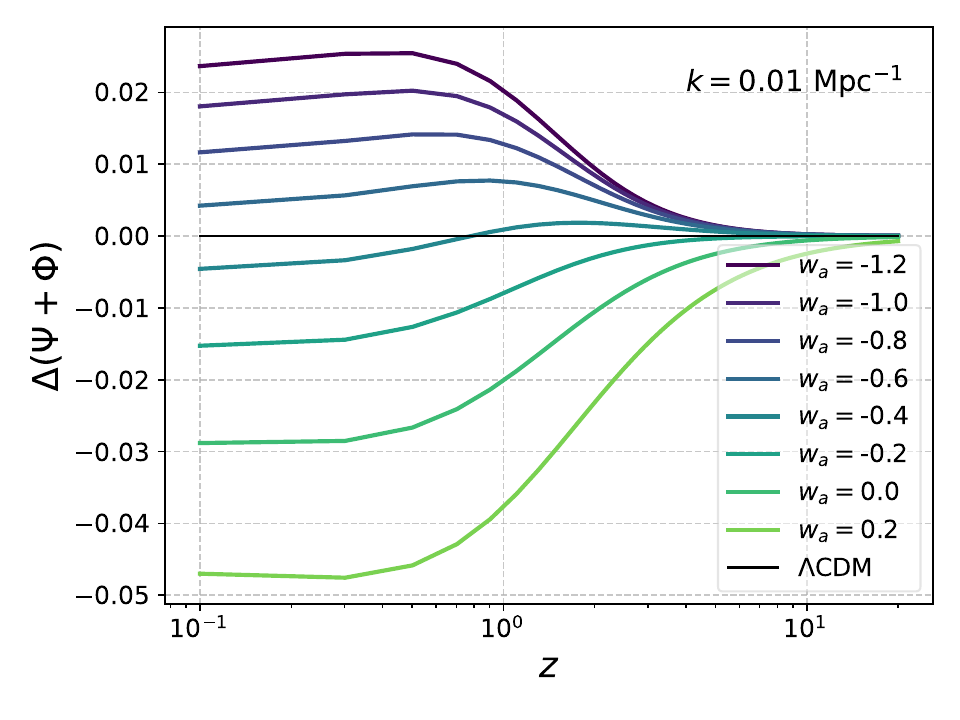}
    \hfill
   \includegraphics[width=0.49\linewidth]{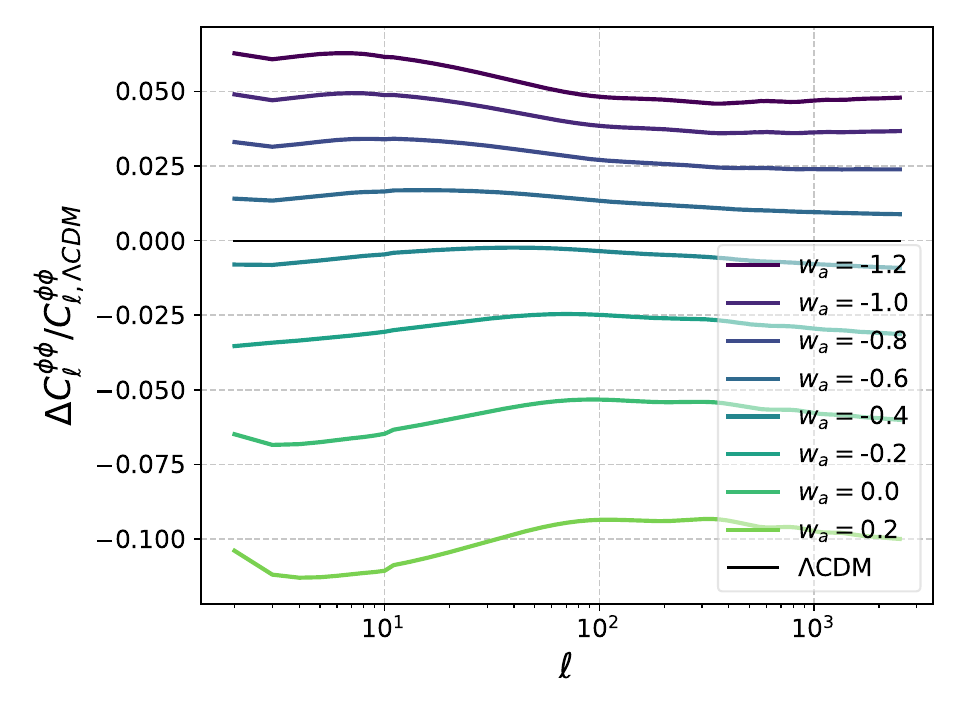}
\caption{The evolution of the Weyl potential as a function of
redshift in $w_0w_a$CDM models compared with $\Lambda$CDM (left),
as well as the corresponding fractional change in the lensing
spectrum $C_{\ell}^{\phi \phi}$ (right). We fix $w_0=-0.85$ and
choose eight different values of $w_a$ in the range $[-1.2,0.2]$.
All other parameters are fixed to the Planck 2018 best-fit values
\cite{Planck:2019nip}.}
    \label{fig:lensing}
\end{figure}

We plot the evolution of the Weyl potential in the $w_0w_a$CDM
model compared with the $\Lambda$CDM model (left) and the
corresponding fractional change in the lensing spectrum
$C_{\ell}^{\phi \phi}$ (right) in Fig.~\ref{fig:lensing}. Inspired
by the DESI results, we fix $w_0=-0.85$ and choose eight different
values of $w_a$ in the range $[-1.2,0.2]$. It can be observed that
only the DE models that cross the phantom divide in the past, i.e.
$w_0+w_a<-1$ preferred by DESI, have the potential to relatively
magnify the Weyl potential, and thus suppress the lensing anomaly.
We can observe a positive correlation between $A_L$ and $w_0+w_a$ in
Fig.~\ref{fig:w0wa_pre}, especially for the Plik+DESI dataset, and
also from the degeneracy direction between $(w_0,w_a)$ and $A_L$
in the right panel of  Fig.~\ref{fig:Al}.

\section{Conclusion}\label{sec:conclusion}

In this paper, we investigate the impact of the lensing anomaly in
Planck data on the nature of DE, using the Planck PR3 and PR4
likelihoods combined with DESI BAO, CMB lensing and Pantheon+
datasets.
The main results are summarized as follows:
\begin{itemize}
\item The DESI BAO combined with Pantheon+ and Planck CMB datasets
do not prefer the evolution of DE ($\Lambda$CDM is $<2\sigma$
consistent), when $A_L$ is allowed to vary\footnote{Two PR4-based
likelihoods slightly differ in their preferences for the evolving
DE when $A_L=1$ is fixed, the CamSpec likelihood maintains the
$>2\sigma$ significant level for evolving DE, while the HiLLiPoP
likelihood slightly weakens it.}. \item The lensing anomaly in the
$\Lambda$CDM model is exacerbated by DESI BAO, which is caused by
the smaller $\Omega_m$ preferred by DESI BAO, relatively
suppressing the lensing potential. However, this effect can be
offset by the shifts in $w_0$ and $w_a$ preferring the evolving
DE.
\end{itemize}
Therefore, the lensing anomaly in Planck data seems have
non-negligible impacts on the nature of DE. In this sense, it is
necessary to further explore the potential systematics in updated
Planck likelihoods and the implications of the lensing anomaly for
DE when one combined new cosmological survey data with CMB.

The nature of DE would also affect our understanding for the very
early universe. Though based on the $\Lambda$CDM model the Planck
collaboration showed the scalar spectral index is $n_s\approx
0.965$ \cite{Planck:2018jri}, $n_s=1$ ($n_s-1\sim {\cal O}
(0.001)$) is favored
\cite{Ye:2020btb,Ye:2021nej,Jiang:2022uyg,Smith:2022hwi,Jiang:2022qlj,Peng:2023bik}
when the early dark energy (EDE) resolution of Hubble tension is
considered\footnote{It seems difficult to ruling out early new
physics by using the sound horizon-free measurement of $H_0$,
e.g.\cite{Jiang:2025ylr}.}, see also
\cite{DiValentino:2018zjj,Giare:2022rvg}, and see recent
Refs.\cite{Kallosh:2022ggf,Braglia:2020bym,Ye:2022efx,Jiang:2023bsz,Braglia:2022phb,DAmico:2021fhz,Giare:2023wzl,Fu:2023tfo,Giare:2024akf}
for its implications for inflation models and discussions. It has
been observed that with DESI data not only the shift towards
$n_s=1$ persists, but also the EDE can suppress the preference for
evolving DE (making $\Lambda$CDM $<2\sigma$ consistent)
\cite{Wang:2024dka}. Thus it can be expected that the results of
DESI about DE might be also affected by different solutions to the
Hubble tension (see
e.g.\cite{Knox:2019rjx,Perivolaropoulos:2021jda,DiValentino:2021izs,Vagnozzi:2023nrq}),
it will be interesting to revisit relevant results in the light of
lensing anomaly.

In all our analyses, we
use the Pantheon+ SN Ia data.
Notably, DES-Y5 SN data \cite{DES:2024jxu}, when combined with
DESI BAO and CMB data, show stronger significant level
($>3\sigma$) for evolving DE compared to Pantheon+. Though it has
been pointed out in Ref.~\cite{Efstathiou:2024xcq} that this
preference is likely a result of systematics in the DES-Y5 sample,
it is worth exploring the impact of lensing anomaly on the nature
of DE using DES-Y5 SN datasets.


\begin{acknowledgments}

We thank Jun-Qian Jiang, Gen Ye for valuable discussion and Eoin \'O Colg\'ain for helpful comments.
This work is supported by NSFC, No.12075246, National Key Research and
Development Program of China, No. 2021YFC2203004, and the
Fundamental Research Funds for the Central Universities.

\end{acknowledgments}

\bibliographystyle{apsrev4-1}
\bibliography{ref}

\appendix

\section{Results of Relevant Parameters}\label{sec:appendix}

\begin{table}[htbp]
    \centering
    \begin{tabular}{|l|c|c|c|}
        \hline
        Parameter & Plik & CamSpec & HiLLiPoP \\
        \hline
        $H_0$                & $67.95(67.91) \pm 0.39$          & $67.74(67.82) \pm 0.37$          & $67.95(67.82) \pm 0.38$ \\
        $100\omega_b$        & $2.246(2.246) \pm 0.013$         & $2.226(2.231) \pm 0.013$         & $2.231(2.231) \pm 0.012$ \\
        $\omega_{\mathrm{cdm}}$ & $0.11866(0.11876) \pm 0.00084$  & $0.11858(0.11847) \pm 0.00081$     & $0.11816(0.11856) \pm 0.00084$ \\
        $10^9A_\mathrm{s}$  & $2.117(2.111)^{+0.026}_{-0.029}$ & $2.108(2.106)^{+0.025}_{-0.029}$ & $2.120(2.111) \pm 0.023$ \\
        $n_\mathrm{s}$      & $0.9683(0.9689) \pm 0.0036$       & $0.9663(0.9657) \pm 0.0036$       & $0.9696(0.9698) \pm 0.0035$ \\
        $\tau_\mathrm{reio}$& $0.0587(0.0570)^{+0.0067}_{-0.0075}$ & $0.0575(0.0559) \pm 0.0072$    & $0.0615(0.0586) \pm 0.0061$ \\
        \hline
        $\chi^2$            & $4203.50$                         & $12404.05$                       & $31994.57$ \\
        \hline
    \end{tabular}
    \caption{The mean (best-fit) $\pm 1\sigma$ errors of relevant parameters and $\chi^2$ values for $\Lambda$CDM model. The datasets are Planck PR3 and PR4 likelihoods, respectively combined with DESI BAO, CMB lensing, and Pantheon+.}
    \label{tab:lcdm}
\end{table}

\begin{table}[htbp]
    \centering
    \begin{tabular}{|l|c|c|c|}
        \hline
        Parameter & Plik & CamSpec & HiLLiPoP \\
        \hline
        $w_{0}$             & $-0.831(-0.826) \pm 0.063$         & $-0.832(-0.838) \pm 0.063$        & $-0.842(-0.856) \pm 0.063$ \\
        $w_{a}$             & $-0.73(-0.73)^{+0.29}_{-0.25}$    & $-0.73(-0.72)^{+0.29}_{-0.25}$    & $-0.64(-0.54)^{+0.30}_{-0.24}$ \\
        \hline
        $H_0$               & $68.05(67.97) \pm 0.72$          & $67.90(67.94) \pm 0.72$          & $67.89(67.78) \pm 0.71$ \\
        $100\omega_b$       & $2.240(2.240) \pm 0.014$         & $2.220(2.215) \pm 0.014$         & $2.226(2.227) \pm 0.013$ \\
        $\omega_{\mathrm{cdm}}$ & $0.11950(0.11943) \pm 0.00099$ & $0.11944(0.11956) \pm 0.00096$     & $0.11880(0.11844) \pm 0.00098$ \\
        $10^9A_\mathrm{s}$ & $2.098(2.098) \pm 0.028$         & $2.088(2.081) \pm 0.028$         & $2.107(2.117) \pm 0.024$ \\
        $n_\mathrm{s}$     & $0.9661(0.9661) \pm 0.0038$       & $0.9641(0.9631) \pm 0.0038$       & $0.9681(0.9696) \pm 0.0037$ \\
        $\tau_\mathrm{reio}$& $0.0544(0.0542) \pm 0.0073$      & $0.0530(0.0510) \pm 0.0072$       & $0.0592(0.0614) \pm 0.0061$ \\
        \hline
        $\chi^2$           & $4195.93$                         & $12396.23$                       & $31988.66$ \\
        \hline
    \end{tabular}
    \caption{The mean (best-fit) $\pm 1\sigma$ errors of relevant parameters and $\chi^2$ values for $w_0w_a$CDM model. The datasets are Planck PR3 and PR4 likelihoods, respectively combined with DESI BAO, CMB lensing, and Pantheon+.}
    \label{tab:w0wacdm}
\end{table}

\begin{table}[htbp]
    \centering
    \begin{tabular}{|l|c|c|c|}
        \hline
        Parameter & Plik & CamSpec & HiLLiPoP \\
        \hline
        $A_L$               & $1.083(1.095)^{+0.033}_{-0.037}$ & $1.075(1.072) \pm 0.034$         & $1.062(1.064) \pm 0.035$ \\
        \hline
        $n_\mathrm{s}$     & $0.9707(0.9711) \pm 0.0037$       & $0.9687(0.9672) \pm 0.0038$       & $0.9716(0.9732) \pm 0.0034$ \\
        $H_0$              & $68.39(68.42) \pm 0.43$          & $68.15(68.08) \pm 0.40$          & $68.26(68.46) \pm 0.39$ \\
        $100\omega_b$      & $2.254(2.251) \pm 0.014$         & $2.235(2.233)^{+0.013}_{-0.012}$ & $2.237(2.243)^{+0.012}_{-0.011}$ \\
        $\omega_{\mathrm{cdm}}$ & $0.11767(0.11760) \pm 0.00093$& $0.11770(0.11784) \pm 0.00088$     & $0.11748(0.11719) \pm 0.00087$ \\
        $\tau_\mathrm{reio}$& $0.0495(0.0514)^{+0.0085}_{-0.0075}$ & $0.0492(0.0493) \pm 0.0083$   & $0.0578(0.0604) \pm 0.0062$ \\
        $10^9A_\mathrm{s}$ & $2.061(2.059) \pm 0.037$         & $2.059(2.057) \pm 0.035$         & $2.085(2.093) \pm 0.030$ \\
        \hline
        $\chi^2$           &  $4197.28$                      &      $12399.81$                  &  $31991.93$   \\
        \hline
    \end{tabular}
    \caption{The mean (best-fit) $\pm 1\sigma$ errors of relevant parameters and $\chi^2$ values for $\Lambda$CDM+$A_L$ model. The datasets are Planck PR3 and PR4 likelihoods, respectively combined with DESI BAO, CMB lensing, and Pantheon+.}
    \label{tab:lcdm+al}
\end{table}

\begin{table}[htbp]
    \centering
    \begin{tabular}{|l|c|c|c|}
        \hline
        Parameter & Plik & CamSpec & HiLLiPoP \\
        \hline
        $A_L$               & $1.066(1.071) \pm 0.041$         & $1.051(1.050) \pm 0.040$         & $1.043(1.037) \pm 0.040$ \\
        $w_{0}$            & $-0.859(-0.862) \pm 0.063$        & $-0.851(-0.866) \pm 0.064$        & $-0.859(-0.856) \pm 0.064$ \\
        $w_{a}$            & $-0.51(-0.45)^{+0.32}_{-0.25}$   & $-0.58(-0.48)^{+0.32}_{-0.26}$   & $-0.52(-0.57)^{+0.30}_{-0.26}$ \\
        \hline
        $H_0$              & $67.92(67.76) \pm 0.73$          & $67.84(67.71) \pm 0.73$          & $67.81(68.20) \pm 0.71$ \\
        $100\omega_b$      & $2.250(2.252) \pm 0.015$         & $2.229(2.231)^{+0.016}_{-0.014}$ & $2.233(2.237)^{+0.014}_{-0.013}$ \\
        $\omega_{\mathrm{cdm}}$ & $0.1182(0.1180) \pm 0.0013$  & $0.1184(0.1183) \pm 0.0012$       & $0.1180(0.1180) \pm 0.0012$ \\
        $10^9A_\mathrm{s}$ & $2.062(2.068)^{+0.037}_{-0.033}$ & $2.061(2.067)^{+0.037}_{-0.032}$ & $2.088(2.092) \pm 0.030$ \\
        $n_\mathrm{s}$     & $0.9695(0.9704) \pm 0.0044$       & $0.9669(0.9662) \pm 0.0043$       & $0.9700(0.9715) \pm 0.0041$ \\
        $\tau_\mathrm{reio}$& $0.0491(0.0520)^{+0.0086}_{-0.0075}$ & $0.0489(0.0504)^{+0.0087}_{-0.0072}$ & $0.0577(0.0581) \pm 0.0062$ \\
        \hline
        $\chi^2$           &   $4192.24$             &    $12394.64$                &  $31986.84$\\
        \hline
    \end{tabular}
    \caption{The mean (best-fit) $\pm 1\sigma$ errors of relevant parameters and $\chi^2$ values for $w_0w_a$CDM+$A_L$ model. The datasets are Planck PR3 and PR4 likelihoods, respectively combined with DESI BAO, CMB lensing, and Pantheon+.}
    \label{tab:w0wacdm+al}
\end{table}

\begin{table}[htbp]
    \centering
    \begin{tabular}{|l|c|c|c|}
        \hline
        Parameter & Plik & CamSpec & HiLLiPoP \\
        \hline
        $A_L$               & $1.062 \pm 0.035$         & $1.054 \pm 0.034$         & $1.042 \pm 0.035$ \\
        \hline
        $H_0$              & $67.84 \pm 0.44$          & $67.62 \pm 0.43$          & $67.74 \pm 0.43$ \\
        $100\omega_b$      & $2.245 \pm 0.014$         & $2.226 \pm 0.014$         & $2.228 \pm 0.013$ \\
        $\omega_{\mathrm{cdm}}$ & $0.11887 \pm 0.00097$& $0.11886 \pm 0.00093$     & $0.11860 \pm 0.00094$ \\
        $10^9A_\mathrm{s}$ & $2.065^{+0.037}_{-0.032}$ & $2.062^{+0.036}_{-0.032}$ & $2.088 \pm 0.030$ \\
        $n_\mathrm{s}$     & $0.9676 \pm 0.0038$       & $0.9657 \pm 0.0038$       & $0.9686 \pm 0.0036$ \\
        $\tau_\mathrm{reio}$& $0.0489^{+0.0086}_{-0.0074}$ & $0.0485^{+0.0083}_{-0.0072}$ & $0.0571 \pm 0.0062$ \\
        \hline
    \end{tabular}
    \caption{The mean $\pm 1\sigma$ errors of relevant parameters for $\Lambda$CDM+$A_L$ model. The datasets are Planck PR3 and PR4 likelihoods, respectively combined with pre-DESI BAO, CMB lensing, and Pantheon+.}
    \label{tab:lcdm+al_pre}
\end{table}

\begin{table}[htbp]
    \centering
    \begin{tabular}{|l|c|c|c|}
        \hline
        Parameter & Plik & CamSpec & HiLLiPoP \\
        \hline
        $A_L$               & $1.067 \pm 0.044$         & $1.052 \pm 0.041$         & $1.046 \pm 0.042$ \\
        $w_{0}$            & $-0.906 \pm 0.062$        & $-0.902 \pm 0.063$        & $-0.905 \pm 0.062$ \\
        $w_{a}$            & $-0.25^{+0.29}_{-0.23}$   & $-0.30^{+0.29}_{-0.24}$   & $-0.26^{+0.28}_{-0.23}$ \\
        \hline
        $H_0$              & $67.16 \pm 0.69$          & $67.06 \pm 0.68$          & $67.07 \pm 0.68$ \\
        $100\omega_b$      & $2.248 \pm 0.016$         & $2.227^{+0.017}_{-0.015}$ & $2.231^{+0.016}_{-0.013}$ \\
        $\omega_{\mathrm{cdm}}$ & $0.1185 \pm 0.0014$  & $0.1187 \pm 0.0013$       & $0.1183^{+0.0012}_{-0.0013}$ \\
        $10^9A_\mathrm{s}$ & $2.063 \pm 0.036$         & $2.062^{+0.035}_{-0.032}$ & $2.087 \pm 0.030$ \\
        $n_\mathrm{s}$     & $0.9686 \pm 0.0047$       & $0.9661 \pm 0.0046$       & $0.9694 \pm 0.0042$ \\
        $\tau_\mathrm{reio}$& $0.0488^{+0.0084}_{-0.0075}$ & $0.0487^{+0.0082}_{-0.0072}$ & $0.0572 \pm 0.0063$ \\
        \hline
    \end{tabular}
    \caption{The mean $\pm 1\sigma$ errors of relevant parameters for $w_0w_a$CDM+$A_L$ model. The datasets are Planck PR3 and PR4 likelihoods, respectively combined with pre-DESI BAO, CMB lensing, and Pantheon+.}
    \label{tab:w0wacdm+al_pre}
\end{table}

\end{document}